\documentclass[%
 reprint,
superscriptaddress,
twocolumn,
 amsmath,amssymb,
 aps,
pra,
]{revtex4-2}

\usepackage{graphicx}
\usepackage{dcolumn}
\usepackage{bm}
\usepackage{xcolor}
\usepackage{float}
\usepackage{soul}
\usepackage{natbib}

\begin{document}

\preprint{APS/123-QED}

\title{Generation of high-order harmonics with tunable photon energy and spectral width using double pulses}

\author{L\'{e}n\'{a}rd Guly\'{a}s Oldal}
\email[email: ]{lenard.gulyas@eli-alps.hu}
\affiliation{ELI-ALPS, ELI-HU Non-Profit Ltd., Wolfgang Sandner utca 3., H-6728 Szeged, Hungary}

\author{Tam\'{a}s Csizmadia}
\affiliation{ELI-ALPS, ELI-HU Non-Profit Ltd., Wolfgang Sandner utca 3., H-6728 Szeged, Hungary}

\author{Peng Ye}
\affiliation{ELI-ALPS, ELI-HU Non-Profit Ltd., Wolfgang Sandner utca 3., H-6728 Szeged, Hungary}

\author{Nandiga Gopalakrishna Harshitha}
\affiliation{ELI-ALPS, ELI-HU Non-Profit Ltd., Wolfgang Sandner utca 3., H-6728 Szeged, Hungary}
\affiliation{Current address:~Friedrich Schiller Universit\"{a}t, Institut f\"{u}r Optik und Quantenelektronik, Jena, Germany}

\author{Amelle Za\"{i}r}
\affiliation{ELI-ALPS, ELI-HU Non-Profit Ltd., Wolfgang Sandner utca 3., H-6728 Szeged, Hungary}
\affiliation{King\textsc{\char13}s College London, Attosecond Physics Group, Department of Physics, WC2R 2LS London, United Kingdom}

\author{Subhendu Kahaly}
\affiliation{ELI-ALPS, ELI-HU Non-Profit Ltd., Wolfgang Sandner utca 3., H-6728 Szeged, Hungary}

\author{Katalin Varj\'{u}}
\affiliation{ELI-ALPS, ELI-HU Non-Profit Ltd., Wolfgang Sandner utca 3., H-6728 Szeged, Hungary}

\author{Mikl\'{o}s F\"{u}le}
\affiliation{ELI-ALPS, ELI-HU Non-Profit Ltd., Wolfgang Sandner utca 3., H-6728 Szeged, Hungary}

\author{Bal\'{a}zs Major}
\affiliation{ELI-ALPS, ELI-HU Non-Profit Ltd., Wolfgang Sandner utca 3., H-6728 Szeged, Hungary}

\date{\today}

\begin{abstract}
This work theoretically investigates high-order harmonic generation in rare gas atoms driven by two temporally delayed ultrashort laser pulses. Apart from their temporal delay, the two pulses are identical. Using a single-atom model of the laser-matter interaction it is shown that the photon energy of the generated harmonics is controllable within the range of one eV -- a bandwidth comparable to the photon energy of the fundamental field -- by varying the time delay between the generating laser pulses. It is also demonstrated that high-order harmonics generated by double pulses have advantageous characteristics, which mimick certain properties of an extreme ultraviolet (XUV) monochromator. With the proposed method, a simpler setup at a much lower cost and comparatively higher spectral yield can be implemented in contrast to other approaches.
\end{abstract}

\maketitle

\section{\label{sec:s1}Introduction}
\vspace{2mm}
High harmonic generation (HHG) in gaseous medium is one of the most widely employed methods to generate coherent radiation in the extreme ultraviolet (XUV) and soft X-ray regions \cite{Akiyama1992,Whalstrom1993,Agostini2004,Subhendu2017}. The process occurring in the interaction of a short laser pulse and a single atom can be described by the semiclassical three-step model \cite{Corkum1993}. First, an electron becomes free by tunneling through the Coulomb potential barrier of the atomic core, which is distorted by the driving laser field. Secondly, the electron wavepacket gains energy from the strong laser field during its motion in the continuum. Finally, it recombines with its parent ion and gets rid of the excess energy by releasing an energetic XUV photon. This process can take place in every half-cycle of the laser field, if the electric field is strong enough to free the electron. Nowadays, HHG is usually driven by state-of-the-art laser systems \cite{Kuhn2017} in order to reach a higher flux of the emitted photons \cite{Klas2018}, a broader photon energy range of the generated harmonics \cite{Sansone2011}, or to obtain an unprecedented time resolution of a few tens of attoseconds \cite{Li2014}; achievements which have already enabled numerous applications \cite{Huillier2003,Reduzzi2015}. These characteristics are strongly influenced by the driving electric field, because strong-field electron dynamics is intricately linked with the time evolution of the driving laser. 
\par
Several theoretical and experimental studies have been carried out using temporally modulated laser fields as seeding sources of HHG to analyze the effect of the temporal evolution of the laser electric field on the HHG process \cite{Ciappina2012,Yavuz2012,Chou2015}. One possible advantage of using a modulated driving field is to extend the cut-off on the single-atom level \cite{Chou2015}. Furthermore, by varying the spectral chirp of the laser pulses, the photon energy of the generated high-order harmonics is finely controllable in a specific spectral range \cite{Lee2001,Kim2003,Chang1998}. There were also successful attempts to generate harmonics by using laser field in combination with a static field, which results in the breaking of the inversion symmetry \cite{Wang1998,Odzak2005,Xiang2009}. Application of such a combined field leads to a multiplateau harmonic structure \cite{Odzak2005,Xiang2009}. The main goal of the above mentioned techniques is to shape the harmonic spectrum by controlling the electron quantum paths during the HHG process via the manipulation of the driving electric field \cite{Merdji2006}. Although special pulse shapes (trapezoidal, square-like, etc.) could provide ways to shape or tune the generated harmonics \cite{Ciappina2012}, nevertheless these pulse shapes are difficult to produce experimentally. 
\par
In addition to the methods that vary the time evolution of the driving field, another experimentally feasible technique to produce laser pulses with unconventional time evolution involves the use of broadband pulses and time delay between the short- and long-wavelength region of their spectrum \cite{Raith2012}. In this way, double pulses can be generated with slightly different central wavelengths. Furthermore, a double-pulse structure can be achieved by interfering two ultrashort laser pulses of the same color. The HHG cut-off can be significantly extended if a carrier envelope phase (CEP) difference of $\pi$ is introduced between the constituting pulses and one of them is shifted by a half-cycle \cite{Perez2009}. Moreover, if the time delay is changed between the pulses, a variety of time evolutions can be introduced when the composing pulses of the produced structure have the same CEP. Although experimental implementation of double-pulse structures is relatively easier, their characterization has inherent issues with most of the well-known pulse characterization techniques, such as Spectral Phase Interferometry for Direct Electric-field Reconstruction (SPIDER) \cite{Iaconis1998} or Frequency-Resolved Optical Gating (FROG) \cite{Trebino1993}. However, procedures have already been developed to measure the temporal shape of double pulses, like the Very Advanced Method for Phase and Intensity Retrieval of E-fields (VAMPIRE) method \cite{Seifert2008} or the recently demonstrated improved Self-Referenced Spectral Interferometry (SRSI) algorithm \cite{Gulyas2019}. 
\par
This work presents theoretical investigations of high harmonic generation in gaseous medium driven by a double-pulse structure. The composing pulses have the same spectrum, spectral phase and absolute CEP, while the control over the temporal evolution of the electric field is reached by changing the delay between them. Holzner et al. \cite{Holzner} proposed and realised a protocol to employ double pulses for HHG attosecond control. In our work we focus on exploring how the double-pulse structure can be successfully employed to tune the harmonics central frequency.
\par
The manuscript is structured as follows: Section \ref{sec:calc} describes the theoretical model and the calculation methods used in this work. The results obtained are summarized in Section \ref{sec:res}, where the impact of the modulated electric field of double pulses on HHG is discussed. Section \ref{sec:conc} presents the main conclusions.

\vspace{5mm}

\section{\label{sec:calc}Theoretical background}
\subsection{Theoretical model}
The interaction of a hydrogen-like atom and a strong electric field can be described by the time-dependent Schr\"{o}dinger equation (TDSE) \cite{Lewenstein1994,Kenichi2010,Nayak2019,Guiseppe2004}. With the introduction of the atomic units and the use of the length gauge, the TDSE has the following form:

\begin{equation}    \label{eq:1}
    i\frac{\partial{\psi(\textbf{r},t)}}{\partial{t}} = \left[ -\frac{1}{2}\nabla^2 + V(\textbf{r}) + \textbf{r}\textbf{E}(t) \right] \psi(\textbf{r},t),
\end{equation}

\noindent 
where $E(t)$ and $V(\textbf{r})$ represent the time-dependent linearly polarized electric field and the atomic potential, respectively. $\psi(\textbf{r},t)$ denotes the electron wave function, where $\textbf{r}$ is the position vector. To determine the solutions of Eq. (\ref{eq:1}) the following assumptions can be made: (i) the contribution of all excited states can be neglected, (ii) the population depletion of the ground state is negligible and (iii) the electron in the continuum can move like a free particle. By applying these approximations (collectively called Strong Field Approximation, SFA \cite{Nayak2019}), the time-dependent dipole moment can be calculated as: 

\begin{equation}    \label{eq:2}
\begin{aligned}
    x(t) & = i\int_{-\infty}^{t}dt^{'} \int d^3\textbf{p} d^*(\bar{p} + \textbf{A}(t))e^{-iS(\textbf{p},t,t^{'})} \\ & \hspace{1.5cm} E(t^{'}) d(\textbf{p} + \textbf{A}(t^{'})) + c.c.,
\end{aligned}
\end{equation}

\noindent
where $\textbf{p}$ represents the canonical momentum. The notation $c.c.$ indicates the complex conjugate of the preceding expression. The expression $\textbf{d}(\textbf{p})$ is the atomic dipole matrix element, which can be expressed as

\begin{equation}    \label{eq:3}
\begin{aligned}
    \textbf{d}(\textbf{p}) = i \left( \frac{1}{\pi\alpha} \right) ^{3/4}\frac{\textbf{p}}{\alpha}e^{-\textbf{p}^2/2\alpha},
\end{aligned}
\end{equation}

\noindent
with $\alpha = I_p$, where $I_p$ is the ionization potential of the target atom. $\textbf{A}(t)$ denotes the vector potential of the driving laser pulse. $S(\textbf{p},t,t^{'})$ is the quasi-classical action, which has the following form

\begin{equation}    \label{eq:4}
    S(\textbf{p},t,t^{'}) = \int_{t^{'}}^{t}dt^{''} \left( \frac{ \left[ \textbf{p} + \textbf{A}(t^{''}) \right]^2 }{2} + I_p \right).
\end{equation}

\noindent
The dipole spectrum $x(\omega_h)$ can be calculated by the Fourier-transform of Eq. (\ref{eq:2}), giving

\begin{equation}    \label{eq:5}
\begin{aligned}
    x(\omega_{h}) & = \int_{-\infty}^{\infty} dt \int_{-\infty}^{t} dt^{'} \int d^{3}\textbf{p} d^{*}(\textbf{p} + \textbf{A}(t))  \\  
    & \hspace{-0.4cm} e^{ \left[ i\omega_{h}t - iS(\textbf{p},t,t^{'}) \right] } E(t^{'}) d(\textbf{p} + \textbf{A}(t^{'})) + c.c.
\end{aligned}    
\end{equation}

\noindent
The expression above is called Lewenstein integral \cite{Lewenstein1994}, which can be used to obtain the spectrum of the generated harmonic field. This is a standard for calculating the nonlinear response of single atoms. The solution of TDSE  includes internal transitions  and electronic structure contribution \cite{Gaarde_2008} that are not included in SFA and usually contribute to lower order harmonics, below the ionisation threshold. SFA has been proven to be an accurate tool for the description of complex field interacting with matter \cite{Le_2016}. In the scope of our results SFA can accurately demonstrate our findings.

\subsection{Investigation of the double-pulse structure}

The total electric field, whose influence on the HHG process is investigated, is expressed as

\begin{equation}    \label{eq:6}
\begin{aligned}
    E(t) & = E_{0}e^{ -2ln2\frac{t^{2}}{\tau^{2}} } e^{i[\omega_{0}t + \varphi_{cep}]} +  \\
    & \hspace{-0.9cm} R \cdot E_{0}e^{-2ln2\frac{(t-\tau_{d})^{2}}{\tau^{2}}} e^{i[\omega_{0}(t - \tau_{d}) + \varphi_{cep}]} + c.c. ,
\end{aligned}    
\end{equation}

\noindent
where $E_0$ is the amplitude, $\tau$ and $\omega_0$ are the transform limited full width at half maximum (FWHM) duration and the central angular frequency of the laser pulses, respectively. R and $\tau_d$ represent the amplitude ratio and the time separation of the composing pulses. $\varphi_{cep}$ is the absolute carrier envelope phase for each pulse. Time separation $\tau_d$ appears as a linear spectral phase term in the frequency domain, which arises from the Fourier shift theorem. In the spectral domain it appears as a phase term of the form $e^{-i \tau_d (\omega - \omega_0)}$, which causes the interference effect in the spectral domain.
\par
\begin{figure}[t]
    \includegraphics[width=1\linewidth]{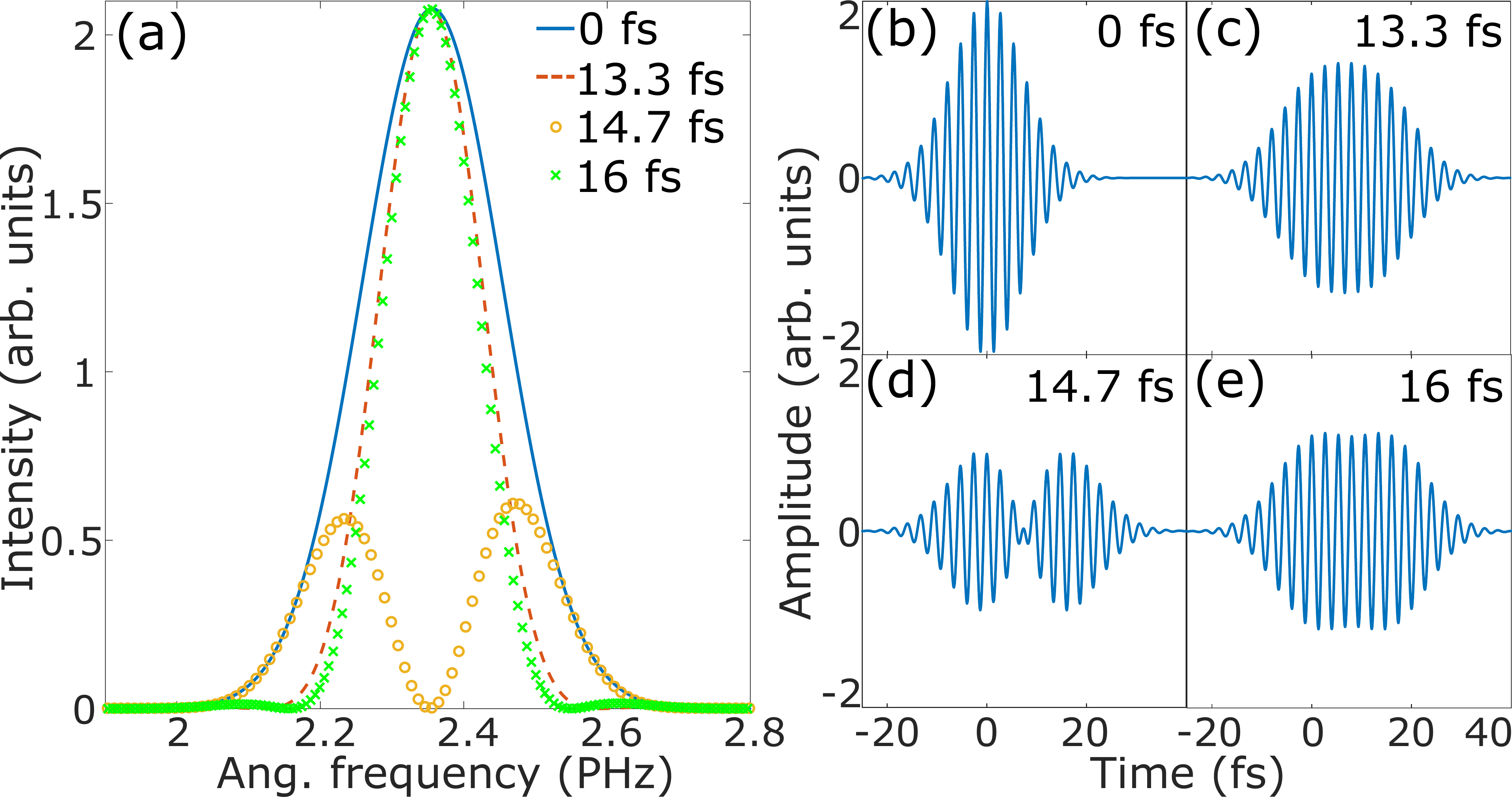}
    \caption{(Color online) (a) The spectra and (b-e) the corresponding temporal amplitude of the double pulses for different time delays: (b) $0~\mathrm{fs}$ (solid blue), (c) $13.3~\mathrm{fs}$ (dashed orange), (d) $14.7~\mathrm{fs}$ (yellow open circles) and (e) $16~\mathrm{fs}$ (green crosses) between the pulses. Their amplitude ratio is chosen to be 1. At delays of $0~\mathrm{fs}$, $13.3~\mathrm{fs}$ and $16~\mathrm{fs}$ the two pulses interfere constructively, while at a delay of $14.7~\mathrm{fs}$ they interfere destructively.}
    \label{fig:pulses}
\end{figure}

Fig. \ref{fig:pulses}(a) depicts the spectra and Fig. \ref{fig:pulses}(b-e) show the temporal profiles of the double-pulse structure for different time delay ($\tau_d$) values. The FWHM duration of both pulses is $12~\mathrm{fs}$ and their central frequency is $2.36~\mathrm{PHz}$ ($800~\mathrm{nm}$ central wavelength). The amplitude ratio between them is 1. The different curves in Fig. \ref{fig:pulses}(a) show how the spectrum changes when the time delay between the two pulses is varied. It reveals that at larger delays spectral valleys appear as a result of the spectral interference. The distance between the spectral minima ($\omega_d$) is directly related to the temporal separation by $\omega_d = 2\pi/\tau_d$. If the temporal amplitude ratio R is 1, the spectral minima reach zero. As the ratio becomes smaller, the spectral amplitude will not reach 0 and the contrast of the interference drops \cite{Paschotta2008}. 
\par
\begin{figure}[t]
    \includegraphics[width=1\linewidth]{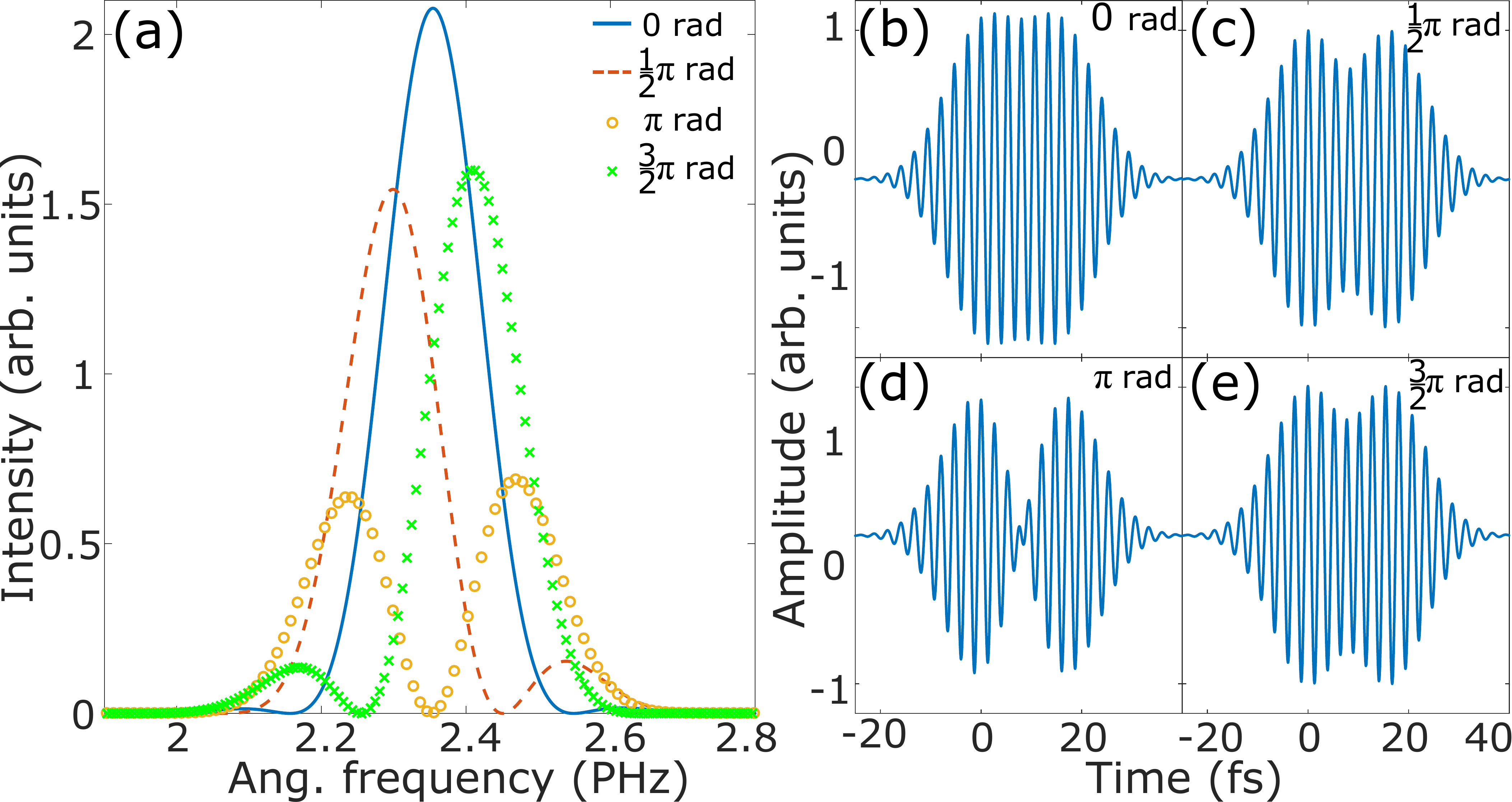}
    \caption{(Color online) (a) The spectral intensity and (b-e) the corresponding temporal evolution at a fixed delay, but (b) at 0, (c) $\pi/2$, (d) $\pi$ and (e) $3\pi/2$ rad relative $\varphi_{cep}$ between the composing pulses of the double-pulse structure. The chosen delay is $\tau_d = 16~\mathrm{fs}$, where the two pulses constructively interfere in case of 0 rad $\varphi_{cep}$. The other parameters are the same as those in Fig. 1.}    \label{fig:cep}
\end{figure}

In fact, the position of the spectral peaks under the envelope in the frequency domain is strongly influenced by the relative $\varphi_{cep}$ difference between the composing pulses. If the delay between the pulses is changed, it also affects the relative $\varphi_{cep}$ between them because they have the same absolute $\varphi_{cep}$. The spectrum and the time evolution of the electric field with different relative CEPs and a fixed time delay between the pulses can be seen in Fig. \ref{fig:cep}. A similar spectral intensity change is observed as in the case of delay variation (Fig. \ref{fig:pulses}(a)). This is a purely linear optical effect caused by the interference of the double pulses. While it is important to consider the effect of $\varphi_{cep}$ in the experimental implementations of double-pulse generation, its influence on harmonic generation is not studied here, because it is analogous to the effect of delay (cf. Fig. \ref{fig:pulses} and Fig. \ref{fig:cep}).

\par
In case of gas HHG (GHHG), on the single atom level, the spectrum of the driving laser is imprinted in the high-order harmonics \cite{Salaries1995,Huillier1991}, thus by modulating the spectrum of the seeding source of the GHHG process, the generated high harmonics are also altered. To simplify and emphasize the main observations and still be experimentally relevant, the propagation effects (such as plasma generation, self-focusing, absorption and dispersion effects) are not included in the simulations. One way to produce similar conditions experimentally is to choose a very thin target, such as a supersonic gas jet \cite{Itatani2004,Itatani2005} and to set up a loose focusing condition, so the gas cell thickness is negligible compared to the Rayleigh length of the laser beam.

\section{\label{sec:res}Results}
\begin{center}
\begin{figure*}[t]
    \includegraphics[width=.9\linewidth]{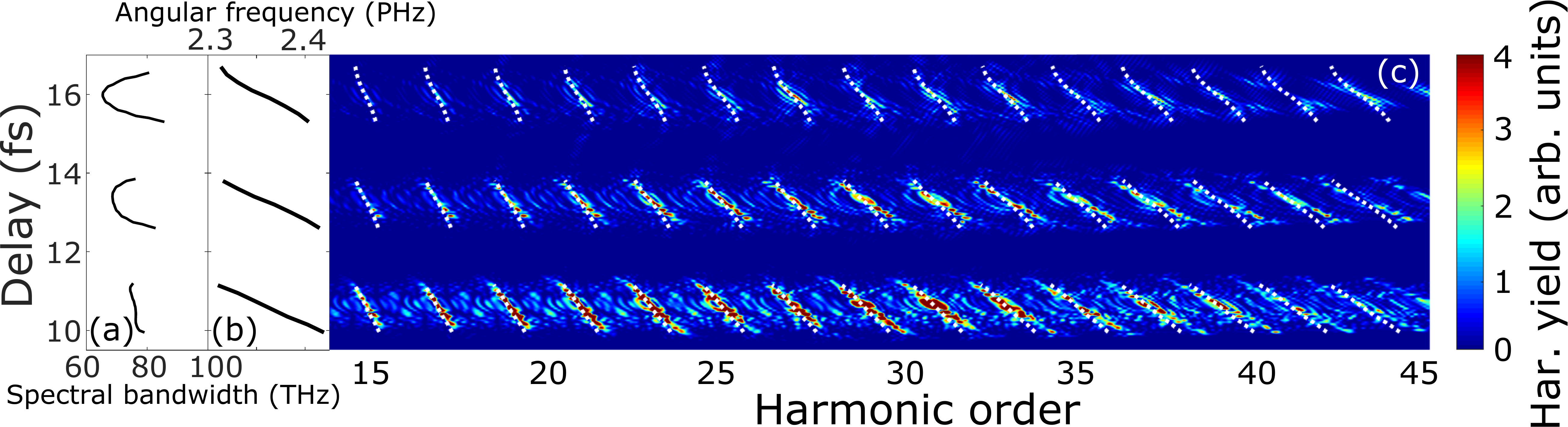}
    \caption{(Color online) (a) Variation of the spectral bandwidth and (b) the central angular frequency, of the driving double pulses changing the time delay between the constituting pulses. The colored map (c) shows the dependence of the high-order harmonic spectrum on the time separation of the two pulses, applying $50~\mathrm{as}$ time resolution. The composing pulses have $12~\mathrm{fs}$ FWHM duration and $800~\mathrm{nm}$ central wavelength. Their amplitude ratio is 1. The target gas is argon. Harmonic order is defined as the higher orders of the central frequency of a single pulse. The harmonic yield (colorbar) is in linear scale. \vspace{-7mm}}
    \label{fig:fullmap}
\end{figure*}
\end{center}

In the simulations, temporal double pulses have been applied as a driving field of the GHHG. The variation of the spectral bandwidth of the driving field with the time separation of the double pulses ($\Delta\omega^2 = \int_{-\infty}^{\infty} (\omega-\omega_0)^2\lvert E(\omega)\lvert^2d\omega/\int_{-\infty}^{\infty} \lvert E(\omega)\lvert^2d\omega$) is presented in Fig. \ref{fig:fullmap}(a). The variance of the central angular frequency ($\omega_0 = \int_{-\infty}^{\infty} \omega\lvert E(\omega)\lvert^2 d\omega/\int_{-\infty}^{\infty} \lvert E(\omega)\lvert^2d\omega$) as the delay changes is represented in Fig. \ref{fig:fullmap}(b). The dotted white curves in Fig. \ref{fig:fullmap}(c) show where each generated harmonic should appear with the central frequency of the generation field at that certain delay ($q \cdot \omega_0$, where $q$ is the harmonic order). The impact of the time separation of the double-pulse structure on the generated high harmonic spectrum is presented in the colored map in Fig. \ref{fig:fullmap}(c). The delay is varied from $9.5$ to $17~\mathrm{fs}$. Both pulses have a Gaussian envelope, while the central wavelength and the FWHM duration are $800~\mathrm{nm}$ and $12~\mathrm{fs}$, respectively. Both pulses have the same $0~\mathrm{rad}$ absolute $\varphi_{cep}$. The target gas is argon, which has an ionization potential ($I_p$) of $15.76~\mathrm{eV}$. As the delay between the pulses is varied, the harmonics at certain delays disappear, while at other delays they show a maximum, which results in distinct bunches in the harmonic structure. This can be explained by the interference of the composing pulses, which interfere constructively at certain delays, and destructively at other

\onecolumngrid

\twocolumngrid

\begin{figure}[b]
    \includegraphics[width=1\linewidth]{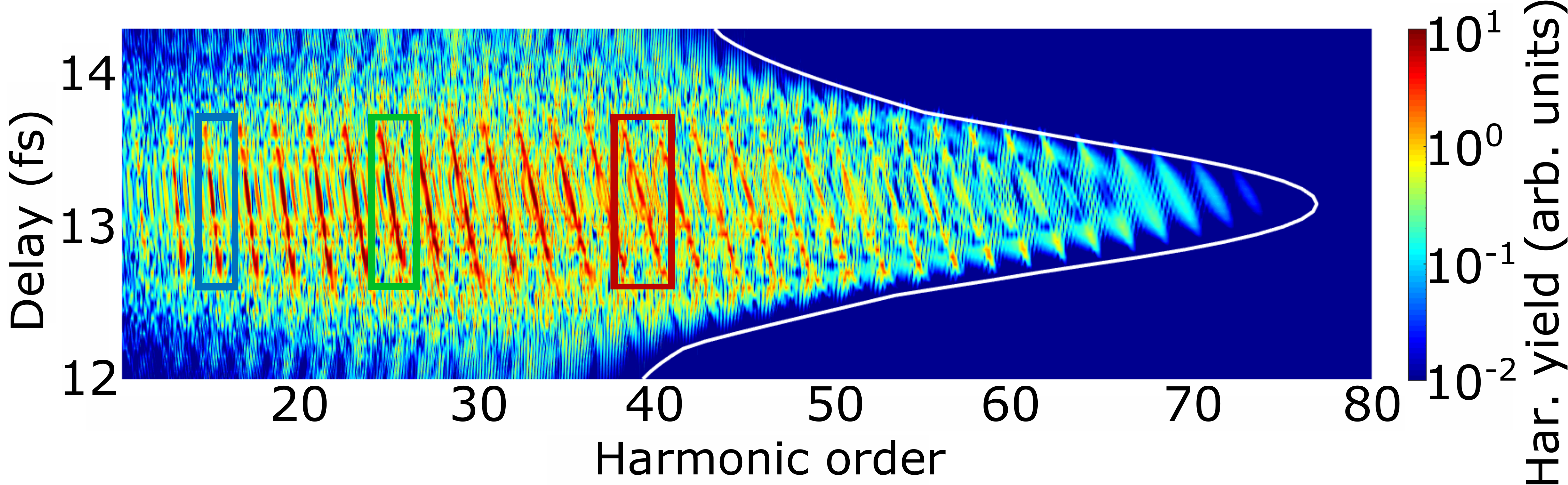}
    \caption{(Color online) Harmonic spectra generated by double pulses. The time separation varies around the $6^{th}$ constructive interference region ($\sim 13.3~\mathrm{fs}$). The other simulation conditions are the same as in Fig. \ref{fig:fullmap}. The white curve represents the calculated cut-off positions $E_c$. The harmonic yield (colorbar) is in logarithmic scale.}
    \label{fig:logspec}
\end{figure}

\noindent
delays (cf. the spectral intensity variation in Fig. \ref{fig:pulses}(a) and \ref{fig:cep}(a)). The appearance of the bunches is determined by the peak-to-peak distance of the driving laser electric field ($T_0 = 2\pi/\omega_0$), which in these simulations equals $2.67~\mathrm{fs}$. The pulses have the same $2.3\cdot10^{14}~\mathrm{\frac{W}{cm^2}}$ peak intensity. This means that when the pulses significantly overlap (around $0~\mathrm{fs}$, $2.67~\mathrm{fs}$, $5.34~\mathrm{fs}$ and $8~\mathrm{fs}$ delay times), the peak intensity is well above the saturation intensity and it is not shown in Fig. \ref{fig:fullmap}. \hspace{1mm} The plot starts from the fourth destructive interference region ($\sim9.5~\mathrm{fs}$ delay). At higher delay values a shift of the individual harmonics can be observed, which results from the variation of the central frequency (cf. Fig. \ref{fig:fullmap}(b)). This allows us to tune the photon energy of the generated harmonics by simply changing the delay between the pulses. The features presented in Fig. \ref{fig:fullmap}(c) have been confirmed by one-dimensional soft-core TDSE calculations. It must be mentioned that in the present simulation the applied amplitude ratio was 1, but the method is robust for dissimilar amplitude ratios too, because the spectral interference contrast is not very sensitive to the amplitude ratio (the contrast change is only 6\% in case of R = 0.7 compared to R = 1). The dominant effect is just a slight blurring of the harmonics, so the method is robust against imperfections in the amplitude ratio.
\par
The harmonic bunch around  $13~\mathrm{fs}$ time separation can be a more interesting position because the tilting phenomenon is clearly visible. It is important to mention that Fig. \ref{fig:fullmap}(c) presents the plateau harmonics only, the cut-off harmonics are not shown. The entire HHG spectra are depicted in Fig. \ref{fig:logspec} for a shorter delay range. The white curve shows the position of the cut-off ($E_c$), which is calculated by the $E_c = I_p + 3.17U_p$ formula. In the expression, $U_p = 9.4\cdot10^{-14}I[\mathrm{W/cm^2}](\lambda[\mathrm{\mu m}])^2$ is the ponderomotive potential, where $I$ and $\lambda$ are the peak intensity and the central wavelength of the driving field \cite{Kenichi2010}, respectively.

\begin{figure}[b]
    \includegraphics[width=1\linewidth]{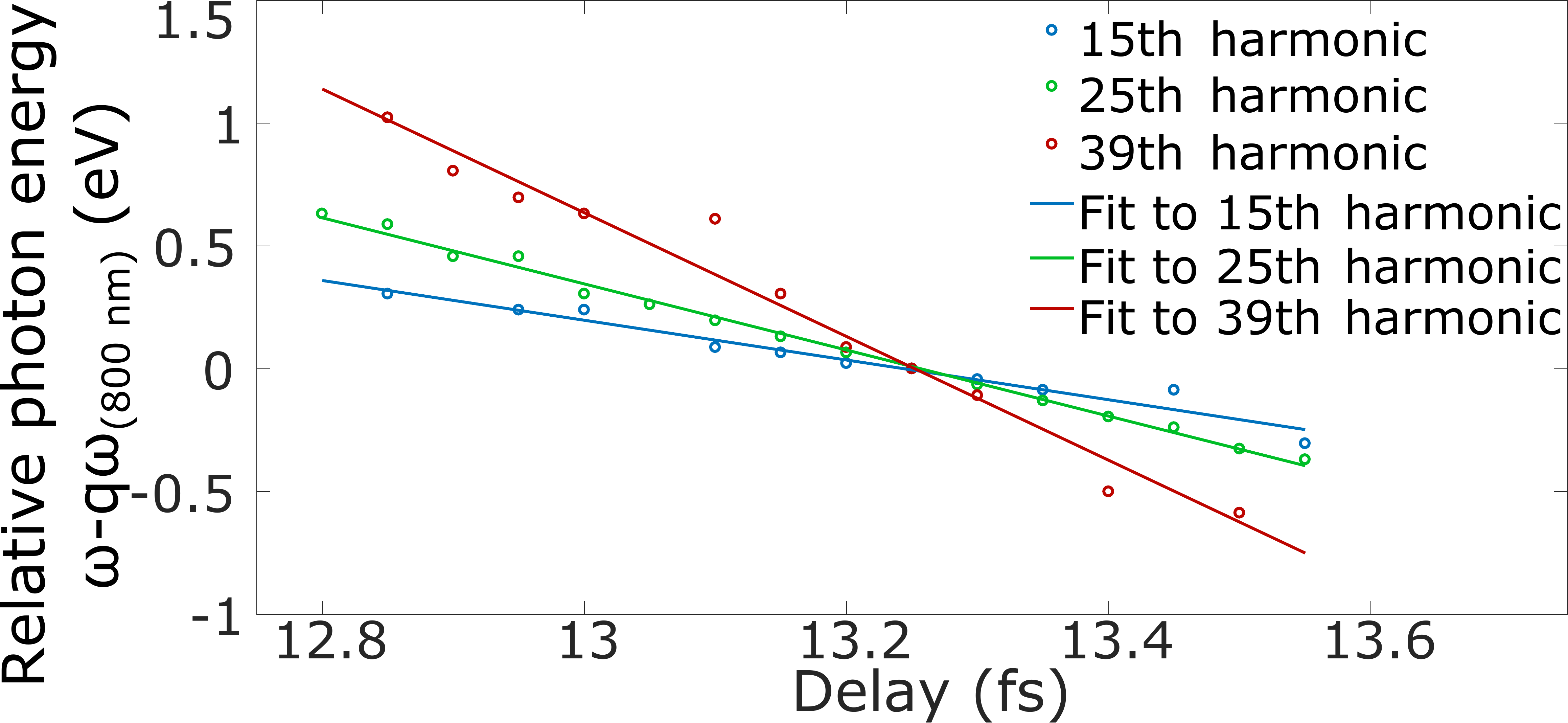}
    \caption{(Color online) The photon energy variation of the $15^{th}$, $25^{th}$ and $39^{th}$ harmonics as the time separation of the double pulses changes. In Fig. \ref{fig:logspec}, these harmonics are marked with appropriately colored rectangles.}
    \label{fig:harmtune}
\end{figure}

\par
To demonstrate the energy tunability range of the individual harmonics, the delay dependence of the spectral position of the $15^{th}$, $25^{th}$ and $39^{th}$ harmonic peaks are presented in Fig. \ref{fig:harmtune}. These harmonics are marked with correspondingly colored rectangles in Fig. \ref{fig:logspec}. It is clearly indicated that as the harmonic order increases, the tilt of the harmonics also increases, so the energy of the higher harmonics can be varied on a wider range. This is a direct consequence of the central frequency change of the generating field (see Fig. \ref{fig:fullmap}(b)). In Fig. \ref{fig:harmtune}, the open circles represent the position of the harmonics obtained from the peak positions in Fig. \ref{fig:logspec}. The solid lines are linear fits to the harmonic peak positions as a function of delay. For better comparison, the harmonics are shifted by their nominal energy value.

\begin{figure}[t]
    \includegraphics[width=1\linewidth]{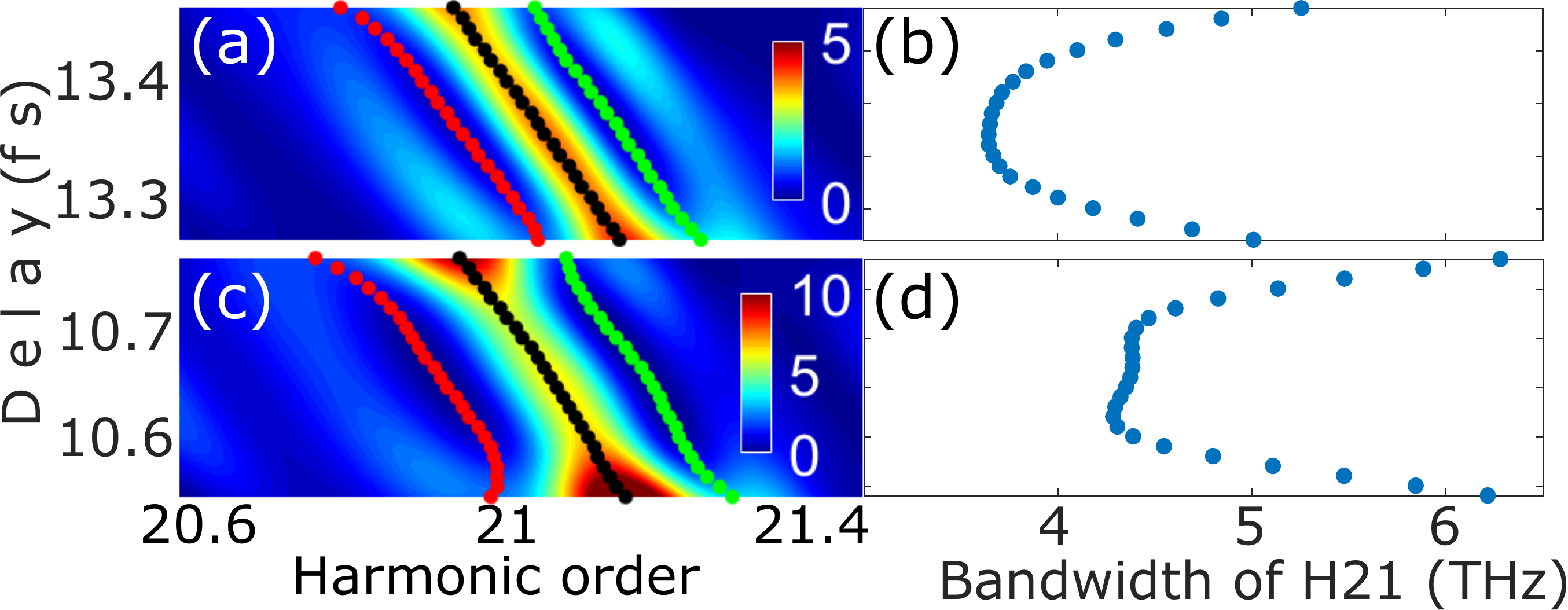}
    \caption{(Color online) Variation of spectral bandwidth of the $21^{st}$ harmonic at two distinct delay regions. (a) and (c) show how the spectrum changes in the vicinity of this harmonic by the delay: the black dots represent the spectral position of the harmonic peak, the red and green dots show the first spectral valley on the low and high photon energy sides of the harmonic peak, respectively. To determine the bandwidth, the region between the red and green dots was used. (b) and (d) demonstrate the spectral bandwidth dependence on the delay.}
    \label{fig:BW}
\end{figure}

\par
Another profitable physical phenomenon that can be observed in Fig. \ref{fig:fullmap}(a) and (c) is the alteration of the width of the generated high-order harmonics. As is known, the harmonics inherit the shape of the original laser spectrum \cite{Salaries1995,Huillier1991}, thus the spectrum of the double-pulse structure appears in the shape of high harmonics. As can be seen in Fig. \ref{fig:fullmap}(a), the spectral width of the driving field slightly decreases as the time separation between the constituting pulses of the double-pulse structure increases. For this reason, moderately thinner harmonics can be generated if double pulses with higher time delays are applied to drive the HHG process. Fig. \ref{fig:BW} displays the spectral bandwidth variation of the $21^{st}$ harmonic as a function of the time separation of double pulses. The spectral region between the red and green dots was used to calculate the bandwidth by the same formula that was used to obtain the bandwidth of the double pulse structure, that is $\Delta\omega^2 = \int_{-\infty}^{\infty} (\omega-\omega_0)^2\lvert E(\omega)\lvert^2d\omega/\int_{-\infty}^{\infty} \lvert E(\omega)\lvert^2d\omega$, with $\omega_0$ being the central frequency of the harmonic peak. As a comparison, the bandwidth of the $21^{st}$ harmonic that is generated by a single Gaussian pulse -- applying the same parameters as used for the composing pulses of the double-pulse structure -- is 9.4 THz, approximately two times the values achievable at the delays presented in Fig. \ref{fig:BW}. It must be also mentioned that the harmonics in the plateau region have the presented spectral evolution, but as they move towards the position of the cut-off the harmonics become more and more blurred, as can be seen in Fig. \ref{fig:logspec}, due to the fewer generated trajectories.
\par
Moreover, the harmonics close to the cut-off region are clearly distinguishable (see Fig. \ref{fig:logspec}). This is closely related to the fact that at bigger time separations (comparable to the pulse duration) the envelope of the driving field begins to resemble a super-Gaussian temporal evolution. For this reason, the total electric field has more half-cycles, which have the same temporal amplitude, as can be seen in Fig. \ref{fig:cep}b, c and e. This results in the emission of attosecond pulses with comparable strength and spectral bandwidth. For this reason, both in the plateau and around the cut-off region the harmonics are sharper compared to the harmonics that are produced by a single Gaussian pulse.

\section{\label{sec:conc}Conclusions}
This paper demonstrates the generation of high-order harmonics with tunable photon energy and spectral width driven by a temporal double-pulse structure. The constituting pulses have a Gaussian temporal profile, having the same spectrum, spectral phase and absolute CEP. Varying the time separation between the composing pulses, the photon energy of the generated harmonics can be varied in an energy range within one eV, which is a bandwidth comparable to the photon energy of the fundamental field. Furthermore, the harmonics are slightly thinner compared to the harmonics that can be produced by a single Gaussian shape electric field. Moreover, when the delayed pulses create a super-Gaussian-like temporal profile, there are multiple, almost identical half-cycles contributing to the emission, thus the cut-off harmonics are as narrow as those in the plateau. For the investigation, SFA is applied to calculate the dipole spectrum. The obtained results are also verified by the numerical solution of the one-dimensional soft-core TDSE. This showed the same effects, which highlights the accuracy of the application of SFA for the present cases.
\par
Harmonics having the presented characteristics can be beneficial for experiments that require a spectrally-tuned single harmonic. Such experiments include angle-resolved photoemission spectroscopy \cite{Mathias2007,Bromberger2015} or lifetime measurements of atomic or molecular excited states \cite{Minda2019,Kolos2016}, which require a narrow wavelength range excitation. Furthermore, photon energy-tunable XUV radiation with a narrow spectral bandwidth is extremely important in many fields such as nanolithograpy \cite{Solak2006}, XUV holography \cite{Bartels2002} or spectroscopy \cite{Arman2012}. Moreover, by combining an XUV multilayer mirror and the photon-energy tunable harmonic generation technique presented here, one can achieve a similar tunability  as with a single-stage XUV monochromator. The advantages of the technique described here over an XUV monochromator include the much lower instrumentation costs, simpler arrangement and consequently easier alignment.
\par
In comparison to the methods that use chirped pulses and are detailed in Refs. [15-17], our procedure gives the possibility to tune the central wavelength of the driving source itself. In contrast to the method presented here, the application of chirped pulses for HHG merely varies the position of the generated harmonics compared to the position of the harmonics produced by transform limited pulses. As another alternative, spectral tuning of the laser field is also possible with wavelength-tunable lasers. Nevertheless, such lasers are relatively rare, and with our method the tuning of the central wavelength can be achieved with any laser source, however only in a narrower range. Moreover, the production of a temporal double-pulse structure can be easily implemented experimentally, for example with a commercially available acousto-optic pulse modulator. Furthermore, our technique can produce a super-Gaussian-like temporal evolution. This quasi top-hat temporal envelope can provide more control over quantum trajectories by controlling other pulse parameters (such as the peak intensity or duration of the pulse), therefore it offers more control over the HHG process in general.

\section{Acknowledgement}
This research was financially supported by the ELI-ALPS project (No. GINOP-2.3.6-15-2015-00001), which was funded by the European Union and co-financed by the European Regional Development Fund. SK would like to acknowledge work from European Cooperation in Science and Technology (COST) Action CA17126 (TUMIEE), supported by COST.


\end{document}